\documentclass[aps,prd,preprint,showpacs,superscriptaddress]{revtex4}
\usepackage{graphicx}
\newcommand{\U}{{\cal U}}
\begin{document}
\draft
\title{Unparticle Physics in Single Top Signals}
\author{A. T. Alan}
\email[ alan\_a@ibu.edu.tr]{}\affiliation{Department of Physics,
Abant Izzet Baysal University, 14280, Bolu, Turkey}
\author{N. K. Pak}
\email[pak@metu.edu.tr]{}
 \affiliation{Department of Physics,
Middle East Technical University, 06531 Ankara, Turkey}
\author{A. Senol}
\email[ senol\_a@ibu.edu.tr]{}
 \affiliation{Department of
Physics, Abant Izzet Baysal University, 14280, Bolu, Turkey}
%\date{\today}

%\date{\today}
\begin{abstract}
We study the single production of top quarks in $e^+e^-, ep$ and
$pp$ collisions in the context of unparticle physics through the
Flavor Violating (FV) unparticle vertices and compute the total
cross sections for single top production as functions of scale
dimension $d_{\U}$. We find that among all, LHC is the most
promising facility to probe the unparticle physics via single top
quark production processes.
\end{abstract}
\maketitle
\section{introduction}
Recently Georgi has proposed a new scheme which is based on the
existence of a nontrivial scale invariant sector at a much higher
scale than that of the Standard Model (SM). He conjectured that this
sector might couple to SM fields via non-renormalizable effective
interactions involving invisible massless objects of fractional
scale dimension, dubbed as unparticles, and thus play a role in low
energy physics \cite{Georgi:2007ek,Georgi:2007si}. This scheme has,
since then, been further developed and studied very extensively,
exploiting it from all phenomenological perspectives \cite{All}.

In this letter we exploit implications of unparticle physics for the
single top production processes. We consider these processes in all
types of colliders, namely International Linear Collider (ILC),
lepton-hadron collider THERA and CERN Large Hadron Collider (LHC)
main parameters such as center of mass energies and integrated
luminosities of which are given in Table \ref{tab3.1b}
\cite{Weiglein:2004hn,Aguilar-Saavedra:2001rg,Abramowicz:2001qt,Branson:2001ak}.
%bizi referan veren ve ILC TDR
For the ILC we consider two options with $\sqrt s=0.5$ and $\sqrt
s=1$ TeV.

In an our earlier paper \cite{Alan:2007ss} we have analyzed the
effects of unparticle physics in the pair production of top quarks
at LHC energies. In those processes there were both Flavor
Conserving (FC) and Flavor Violating (FV) contributions, the FC ones
having counterparts in the framework of SM, and there were some kind
of competition between these two types of contributions. However,
the single production processes, on the other hand, always manifest
themselves via FV vertices. Because in these processes the top
quarks are always accompanied by an antiquark with a different
flavor, the actual type of which depending on the relevant
production channel. As these FV vertices do not exist in the SM,
none of these reactions considered here have SM counterparts, and
all are purely unparticle processes.

The propagators and effective interactions of scalar, vector and
tensor unparticles were already given in several references
\cite{Georgi:2007ek,Georgi:2007si,Alan:2007ss,Cheung:2007ap,Chen:2007qr},
therefore we are not going to present them here once more and refer
the reader to those references.

\section{Differential cross sections for single top productions in unparticle physics}
In this section we will present the differential cross sections by
considering the contributions of all three types of unparticles,
vector, tensor and scalar, for the single top productions through
the processes $e^+e^-\rightarrow t\bar q$, $ep\rightarrow et+X$ and
$pp\rightarrow t(\bar q, g, \U)+X$.
\subsection{$e^+e^-\rightarrow t\bar q$}
Single top production in $e^+e^-$ collisions proceeds via s-channel
unparticle exchange only. The differential cross sections for
vector, tensor and scalar unparticle contributions are given as
\begin{eqnarray}
\frac{d\hat\sigma_V}{d\hat t}&=&\frac{3|\tilde A_V|^2}{8\pi\hat
s^2(\hat s)^{4-2d_{\mathcal{U}}}}\Big[(c_v^4+c_a^4)(\hat t^2+(\hat
s+\hat t)^2-(2\hat t+\hat s)m^2)\nonumber\\&&+2c_v^2c_a^2(2\hat
t^2+3\hat s(\hat s+2\hat t)-(3\hat s+2\hat t)m^2)\Big]
\end{eqnarray}
where \begin{eqnarray} \tilde
A_V=\frac{\lambda_1^2A_{d_{\mathcal{U}}}e^{-i\pi
d_{\U}}}{2\sin(d_{\mathcal{U}}\pi)\Lambda^{2(d_{\mathcal{U}}-1)}},
\end{eqnarray}
\begin{eqnarray}
 \frac{d\hat\sigma_{T}}{d\hat
t}&=&\frac{3|\tilde A_T|^2}{2\pi\hat s^2(\hat
s)^{4-2d_{\mathcal{U}}}}\Big[-(8\hat t+\hat s)m^6+(3\hat s^2+26\hat
t\hat s+40\hat
t^2)m^4\nonumber\\
&-&(3\hat s^3+28\hat t\hat s^2+82\hat t^2\hat s+64\hat t^3)m^2 +\hat
s^4+32\hat t^4+64\hat s\hat t^3+42\hat s^2\hat t^2+10s^3\hat
t)\Big]
\end{eqnarray}
where \begin{eqnarray} \tilde
A_T=\frac{\lambda_2^2A_{d_{\mathcal{U}}}e^{-i\pi
d_{\U}}}{32\sin(d_{\mathcal{U}}\pi)\Lambda^{2d_{\mathcal{U}}}},
\end{eqnarray}
\begin{eqnarray}
\frac{d\hat\sigma_{S}}{d\hat t}&=&\frac{3|\tilde A_S|^2}{4\pi\hat
s^2(\hat s)^{4-2d_{\mathcal{U}}}}\Big[\hat s(\hat s-m^2)\Big]
\end{eqnarray}
where \begin{eqnarray} \tilde
A_S=\frac{\lambda_0^2}{\lambda_1^2}\tilde A_V
\end{eqnarray}
\subsection{$ep\rightarrow
et+X$} Production of single top quarks in $ep$ collisions through
the sub-processes  $eq\rightarrow et$ proceeds via t-channel
unparticle exchange only. The differential cross sections in this
case are
\begin{eqnarray}
\frac{d\hat\sigma_V}{d\hat t}&=&\frac{|\tilde A_V|^2}{8\pi\hat
s^2(-\hat t)^{4-2d_{\mathcal{U}}}}\Big[(c_v^4+c_a^4)(\hat s^2+(\hat
s+\hat t)^2-(2\hat s+\hat t)m^2)\nonumber\\&&-2c_v^2c_a^2(\hat
t^2+2\hat s(\hat t-\hat s)+(2\hat s-\hat t)m^2)\Big]
\end{eqnarray}

\begin{eqnarray}
 \frac{d\hat\sigma_{T}}{d\hat
t}&=&\frac{|\tilde A_T|^2}{2\pi\hat s^2(-\hat
t)^{4-2d_{\mathcal{U}}}}\Big[(-8\hat s-\hat t)m^6+(40\hat s^2+26\hat
t\hat s+3\hat
t^2)m^4\nonumber\\
&-&(64\hat s^3+82\hat t\hat s^2+28\hat t^2\hat s+3\hat t^3)m^2
+32\hat s^4+\hat t^4+10\hat s\hat t^3+42\hat s^2\hat t^2+64\hat
s^3\hat t\Big]
\end{eqnarray}

\begin{eqnarray}
\frac{d\hat\sigma_{S}}{d\hat t}&=&\frac{|\tilde A_S|^2}{4\pi\hat
s^2(-\hat t)^{4-2d_{\mathcal{U}}}}\Big[-\hat t(\hat s+\hat t)\Big]
\end{eqnarray}

\subsection{$pp\rightarrow t(\bar q, g, \U)+X$}
In $pp$ collisions at the LHC there are a rich variety of mechanisms
giving rise to single top productions as compared to the ones
discussed above. There are three types of sub-processes:\\
$i$) $q\bar q \rightarrow t\bar q$ \\
$ii$) $gg\rightarrow t\bar q$\\
$iii$) $qg\rightarrow tg, t\U$

For the first type of processes we have both s- and t-channel
unparticle contributions. The differential cross section for each
types of unparticles are given as
\begin{eqnarray}
\frac{d\hat\sigma_{V}}{d\hat t}&=&\frac{|\tilde A_V|^2}{8\pi\hat
s^2}\Bigg\{\frac{1}{(\hat s)^{4-2d_{\U}}}\Big[(c_v^4+c_a^4)((\hat
s+\hat t)^2+\hat t^2-m^2(\hat s+2\hat
t)\nonumber\\&+&2c_v^2c_a^2(3\hat s^2+6\hat
s\hat t+2\hat t^2-m^2(3\hat s+2\hat t)))\Big]\nonumber\\
&+&\frac{1}{(-\hat t)^{4-2d_{\U}}}\Big[(c_v^4+c_a^4)((\hat s+\hat
t)^2+\hat s^2-m^2(2\hat s+\hat t)\nonumber\\&+&2c_v^2c_a^2(2\hat
s^2+6\hat
s\hat t+3\hat t^2-m^2(2\hat s+3\hat t)))\Big]\nonumber\\
&+&\frac{2\cos d_{\U}\pi}{3(\hat s)^{2-d_{\U}}(-\hat
t)^{2-d_{\U}}}\Big[(c_v^4+6c_a^2c_v^2+c_a^4)(m^2-\hat s-\hat t)(\hat
s+\hat t))\Big]\Bigg\}
\end{eqnarray}

\begin{eqnarray}
\frac{d\hat\sigma_{T}}{d\hat t}&=&\frac{|\tilde A_T|^2}{2\pi\hat
s^2}\Bigg\{\frac{1}{(\hat
s)^{4-2d_{\U}}}\Big[\hat s^4+10\hat s^3\hat t+42\hat s^2\hat t^2+64\hat s\hat t^3+32\hat t^4-m^6(\hat s+8\hat t)\nonumber\\
&+&m^4(3\hat s^2+26\hat s\hat t+40\hat t^2)-m^2(3\hat s^3+28\hat s^2\hat t+82\hat s\hat t^2+64\hat t^3)\Big]\nonumber\\
&+&\frac{1}{(-\hat t
)^{4-2d_{\U}}}\Big[32\hat s^4+64\hat s^3\hat t+42\hat s^2\hat t^2+10\hat s\hat t^3+\hat t^4-m^6(8\hat s+\hat t)\nonumber\\
&+&m^4(40\hat s^2+26\hat s\hat t+3\hat t^2)-m^2(64\hat s^3+82\hat s^2\hat t+28\hat s\hat t^2+3\hat t^3)\Big]\nonumber\\
&+&\frac{\cos d_{\U}\pi}{3(\hat s)^{2-d_{\U}}(-\hat
t)^{2-d_{\U}}}\Big[2m^6(\hat s+\hat t)-(\hat s+\hat t)^2(4\hat s^2+17\hat s\hat t+4\hat t^2)\nonumber\\
&-&m^4(8\hat s^2+21\hat s\hat t+8\hat t^2) +2m^2(5\hat s^3+22\hat
s^2\hat t+22\hat s\hat t^2+5\hat t^3)\Big]\Bigg\}
\end{eqnarray}
\begin{eqnarray}
\frac{d\hat\sigma_{S}}{d\hat t}&=&\frac{|\tilde A_S|^2}{4\pi\hat
s^2}\Bigg\{\frac{1}{(\hat s)^{4-2d_{\U}}}\Big[(m^2-\hat t)(\hat
s+\hat t-m^2)\Big] +\frac{1}{(-\hat t
)^{4-2d_{\U}}}\Big[\hat t(\hat t-m^2)\Big]\nonumber\\
&-&\frac{\cos d_{\U}\pi}{3(\hat s)^{2-d_{\U}}(-\hat
t)^{2-d_{\U}}}(\hat s\hat t)\Bigg\}
\end{eqnarray}

The second type is the gluon fusion which proceeds via s-channel
scalar and tensor unparticle exchanges only, and the differential
cross sections are given as:

\begin{eqnarray}
\frac{d\hat\sigma_{T}}{d\hat t}&=&\frac{3|\tilde A_T|^2}{\pi\hat s^2
(\hat s)^{4-2d_{\U}}} \Big[-\hat t((\hat s+2\hat t)m^4-(2\hat
s^2+5\hat t\hat s+4\hat t^2)m^2\nonumber\\&+&\hat s^2(\hat s+3\hat
t)+2\hat t^2(\hat t+2\hat s))\Big]
\end{eqnarray}
\begin{eqnarray}
\frac{d\hat\sigma_{S}}{d\hat t}&=&\frac{3|\tilde
A_S^g|^2}{256\pi\hat s^2(\hat s)^{4-2d_{\U}}}\Big[\hat s^2(\hat
s-m^2)\Big]
\end{eqnarray}
where
\begin{eqnarray}
\tilde A_S^g=\frac{2\lambda_0^2A_{d_{\mathcal{U}}}e^{-i\pi
d_{\U}}}{\sin(d_{\mathcal{U}}\pi)\Lambda^{2d_{\mathcal{U}}-1}}
\end{eqnarray}

The last group of processes involve associated production of top
quarks with gluons which proceed via t-channel scalar and tensor
exchanges, as well as the rather peculiar process of associated
productions of $t$ quarks with scalar and tensor unparticle through
s- and u-channel $q$ (initial quark) and $t$ quark exchanges,
respectively. Here only scalar and tensor unparticles are produced
in association with the top quark.

Differential cross sections for the subprocess $qg\rightarrow tg$
are
\begin{eqnarray}
\frac{d\hat\sigma_{T}}{d\hat t}&=&\frac{8|\tilde A_T|^2}{\pi\hat s^2
(-\hat t)^{4-2d_{\U}}} \Big[\hat s((2\hat s+\hat t)m^4-(4\hat
s^2+5\hat t\hat s+2\hat t^2)m^2\nonumber\\&+&2\hat s^2(\hat s+2\hat
t)+\hat t^2(\hat t+3\hat s))\Big]
\end{eqnarray}
\begin{eqnarray}
\frac{d\hat\sigma_{S}}{d\hat t}&=&\frac{|\tilde A_S^g|^2}{32\pi\hat
s^2(-\hat t)^{4-2d_{\U}}}\Big[\hat t^2(m^2-\hat t)\Big]
\end{eqnarray}

and finally the differential cross sections for the subprocesses
$qg\rightarrow t\U$

\begin{eqnarray}
\frac{d\hat\sigma_{T}}{d\hat t}&=&\frac{\lambda_2^2\alpha_s}{192\hat
s^2 \Lambda^{2d_{\U}}} \Big[-\frac{3\hat tm^2+4\hat s^2+5\hat s\hat
t}{\hat s}+\frac{2}{(\hat u-m^2)^2}(4m^6-4(2\hat s+\hat
t)m^4\nonumber\\&+&(2\hat t^2+\hat t \hat s-\hat s^2)m^2+2(\hat
s-\hat t)(\hat s+\hat t)^2)+\frac{2}{\hat s(\hat u-m^2)}(4m^6-(7\hat
s+5\hat t)m^4\nonumber\\&+&(\hat s^2+6\hat t\hat s+\hat
t^2)m^2-2\hat s\hat t(\hat s+\hat t))\Big]
\end{eqnarray}
\begin{eqnarray}
\frac{d\hat\sigma_{S}}{d\hat t}&=&\frac{\lambda_0^2\alpha_s}{12\hat
s^2\Lambda^{2d_{\U}-2}}\Big[\frac{\hat s+\hat t}{\hat
s}-\frac{2m^4-(\hat s+\hat t)(2m^2+\hat s)}{(\hat
u-m^2)^2}-\frac{2m^4-2(\hat s+\hat t)(m^2+\hat s)}{\hat s(\hat
u-m^2)}\Big]
\end{eqnarray}
\begin{table}[h]
\caption{The main parameters of future colliders}\label{tab3.1b}
\centering \vspace{0.5cm}
\begin{tabular}{l c c c c } \hline\hline
$e^+e^-$ (ILC)&$E_{e^+}$ (TeV)&$E_{e^-}$ (TeV)&$\sqrt s$ (TeV)
&$L_{int}$($pb^{-1}$) \\\hline  &0.25 (0.5)&0.25 (0.5)&0.5
(1)&$10^4$ \\\hline\hline $ep$ (THERA) &$E_e$ (TeV)&$E_p$
(TeV)&$\sqrt s$ (TeV) &$L_{int}$($pb^{-1}$)
\\\hline
 &0.25&1&1&40
 \\\hline\hline $pp$ (LHC)&$E_{p}$ (TeV)&$E_{p}$
(TeV)&$\sqrt s$ (TeV) &$L_{int}$($pb^{-1}$) \\\hline
&7&7&14&$10^5$\\\hline\hline
\end{tabular}
\end{table}

\section{Numerical Analysis}
\begin{figure}
  \includegraphics[width=12cm]{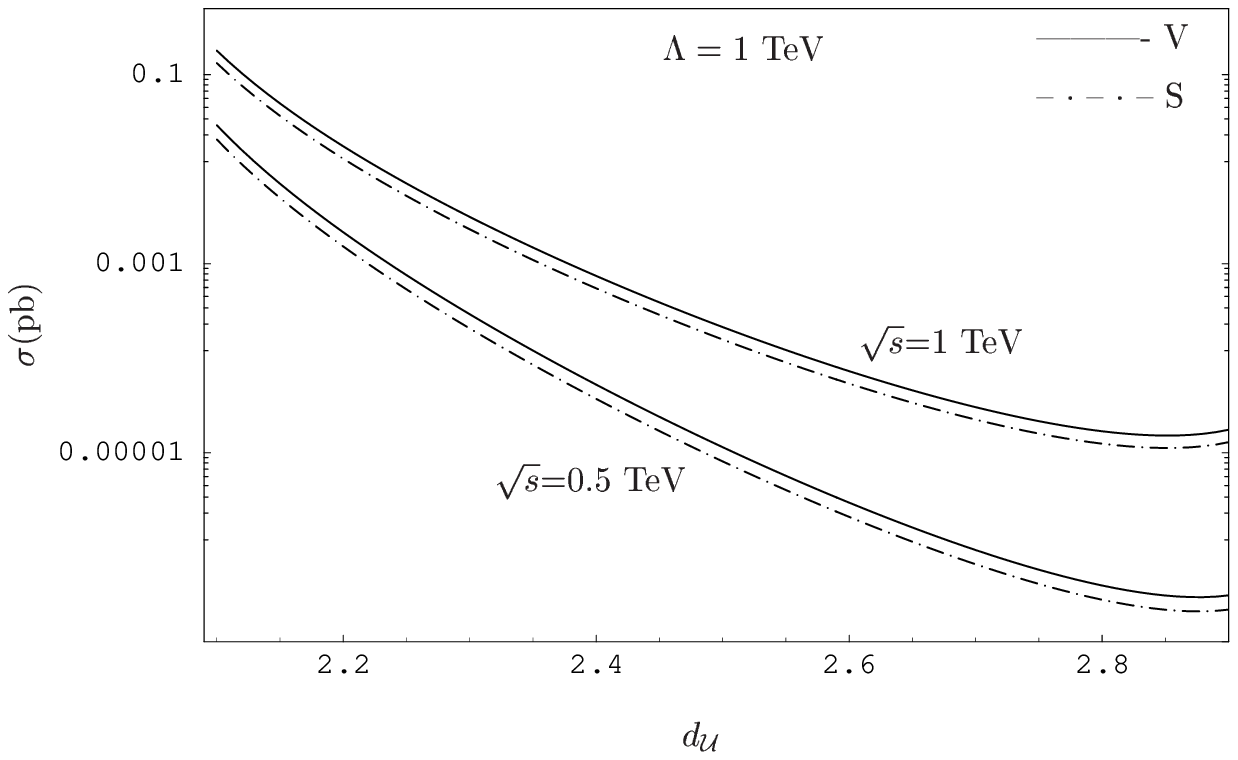}\\
  \caption{Total cross section in pb originating from the reaction $e^+e^-\rightarrow \bar q t$ for $\Lambda$= 1 TeV, $\lambda_0=\lambda_1=1$,
 and $c_v=c_a=1$ at ILC($\sqrt s$=0.5 TeV and $\sqrt s$=1 TeV)}\label{fig1}
\end{figure}

In Fig.~\ref{fig1}, we plot the $d_{\U}$ dependencies of cross
sections for both vector and scalar unparticle contributions in the
interval $2<d_{\U}<3$ for the sake of practical advantage of
depicting both contributions in the same figure. Namely there is a
constraint on $d_{\U}$ ,$d_{\U}>2$, for vector unparticle, in the
case of FV reactions \cite{Choudhury:2007js}, but non for the scalar
mediator. In the figure both options of ILC were considered for the
$\Lambda$=1 TeV value of the mass scale. We see that for a very
narrow interval of $2<d_{\U}<2.12$, effects of vector and scalar
unparticles can be observed at ILC with $\sqrt s= 500$ GeV and with
an integrated luminosity, $L=10^4$ pb$^{-1}$ assuming an
observability limit of number of the single top events to be about
one hundred. Furthermore one gets larger number of events for scalar
unparticles by considering the region $1<d_{\U}<2$, as can easily be
seen by extrapolating the dashed lines in Fig.1. For tensor
unparticles the relevant interval is $3<d_{\U}<4$. In this range the
cross sections are too small. To give an example
$\sigma=1.5\times10^{-8}$ pb for $d_{\U}$=3.1 at $\sqrt s=500$ GeV.

\begin{figure}
  \includegraphics[width=12cm]{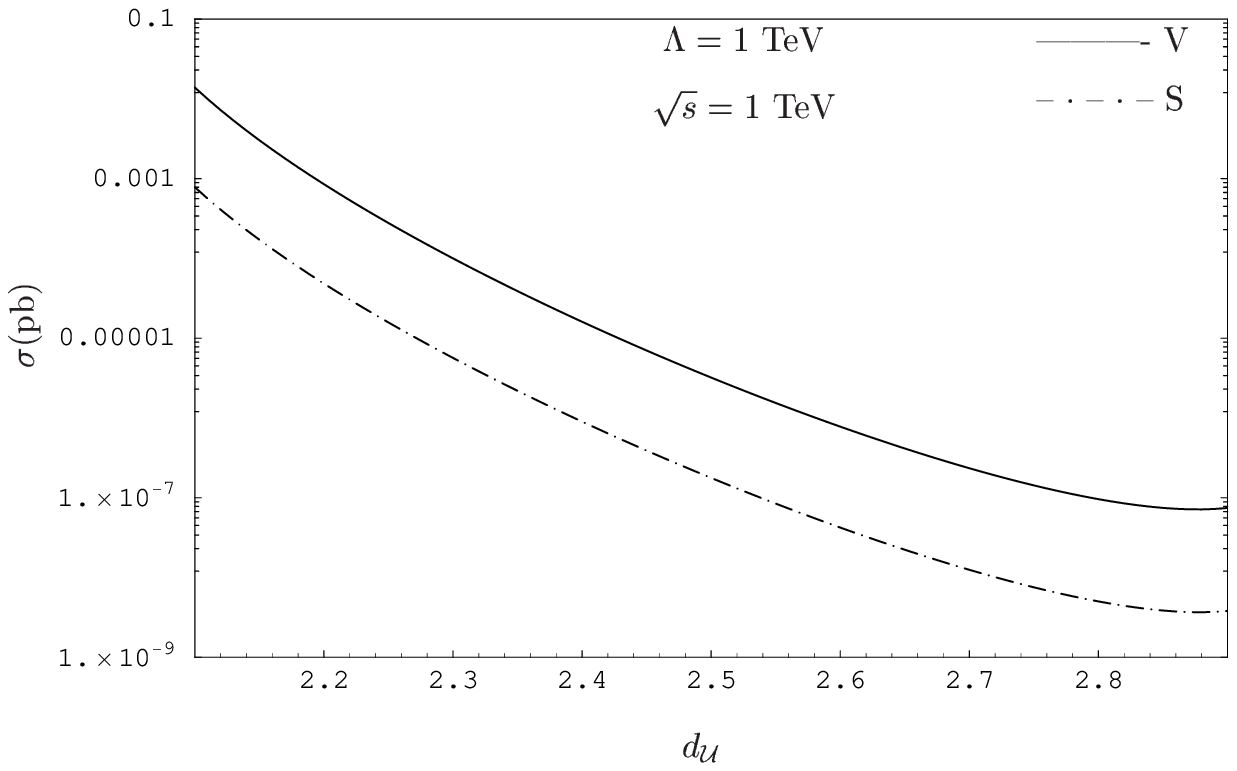}\\
  \caption{Total cross section in pb originating from the reaction $e q_i\rightarrow e t$ for $\Lambda$= 1 TeV, $\lambda_0=\lambda_1=1$,
 and $c_v=c_a=1$ at THERA}\label{fig2}
\end{figure}

In Fig.~\ref{fig2} we have plotted the cross sections for the vector
and scalar cases, originating from the reaction $eq_i\rightarrow et$
($q_i=u,c$) for $\Lambda$=1 TeV, at THERA with $\sqrt s=1$ TeV.
Furthermore, tensor contributions are 8 order smaller than that of
scalar case. Taking into account the fact that integrated luminosity
$L=40$ pb$^{-1}$, then its clear that THERA will not be a suitable
platform to probe unparticle physics. In numerical calculations we
used CTEQ5 parton distributions \cite{CTQ5}.

\begin{figure}
  \includegraphics[width=12cm]{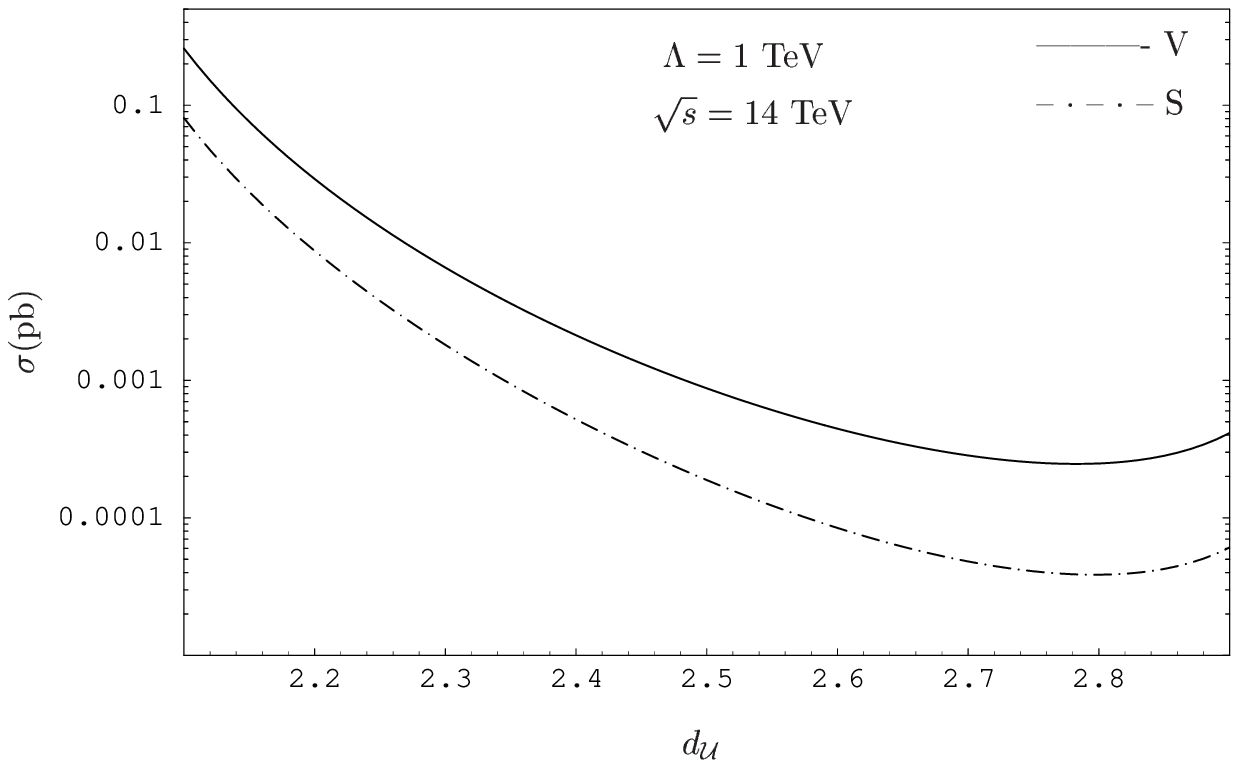}\\
  \caption{Total cross section in pb for the reaction $pp\rightarrow t\bar q_i+X$ ($q_i=u,c$) for $\Lambda$= 1 TeV, $\lambda_0=\lambda_1=1$,
 and $c_v=c_a=1$ at LHC}\label{fig3}
\end{figure}

\begin{figure}
  \includegraphics[width=12cm]{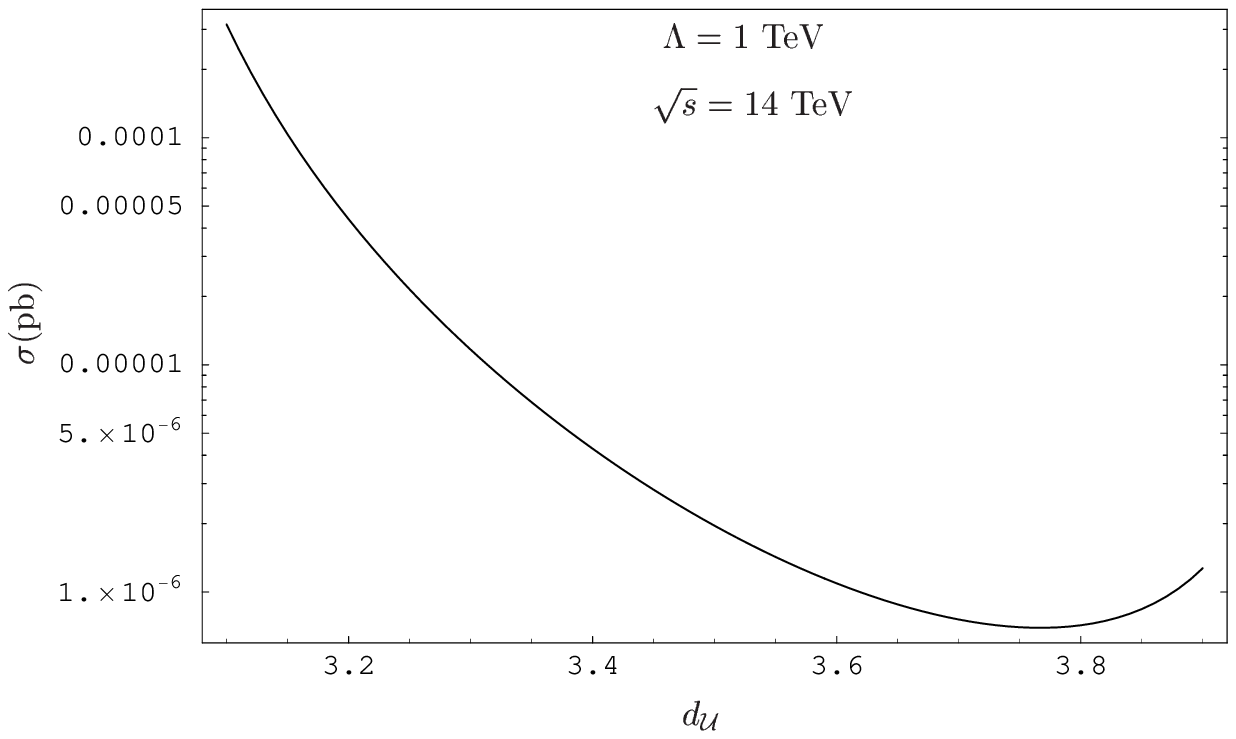}\\
  \caption{Total cross section in pb for the reaction $pp\rightarrow t\bar q_i+X$ ($q_i=u,c$)
   for $\Lambda$= 1 TeV, $\lambda_2=1$,
 at LHC through tensor unparticle}\label{fig4}
\end{figure}

\begin{figure}
  \includegraphics[width=12cm]{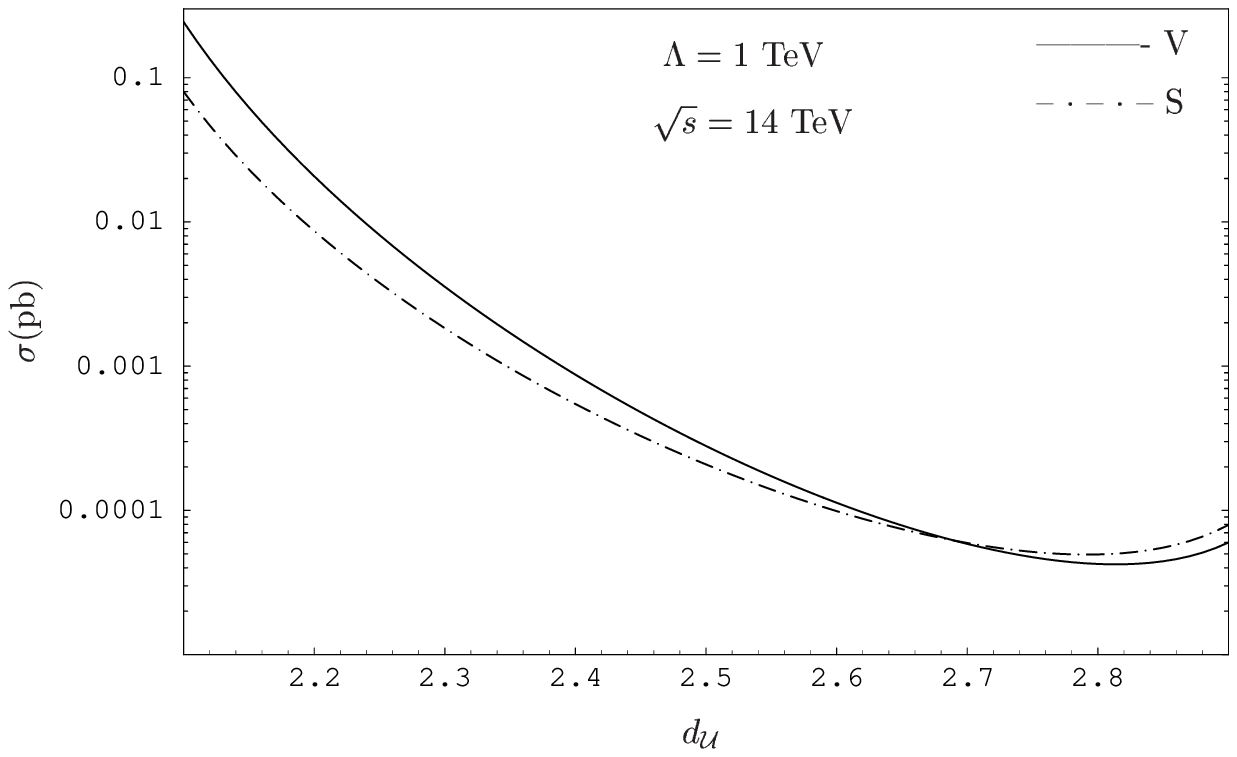}\\
  \caption{Total cross section in pb originating from the reaction $q_i \bar q_j\rightarrow t\bar q_j$ ($q_j=d,b,s$)
  for $\Lambda$= 1 TeV, $\lambda_0=\lambda_1=1$,
 and $c_v=c_a=1$ at LHC}\label{fig5}
\end{figure}

\begin{figure}
  \includegraphics[width=12cm]{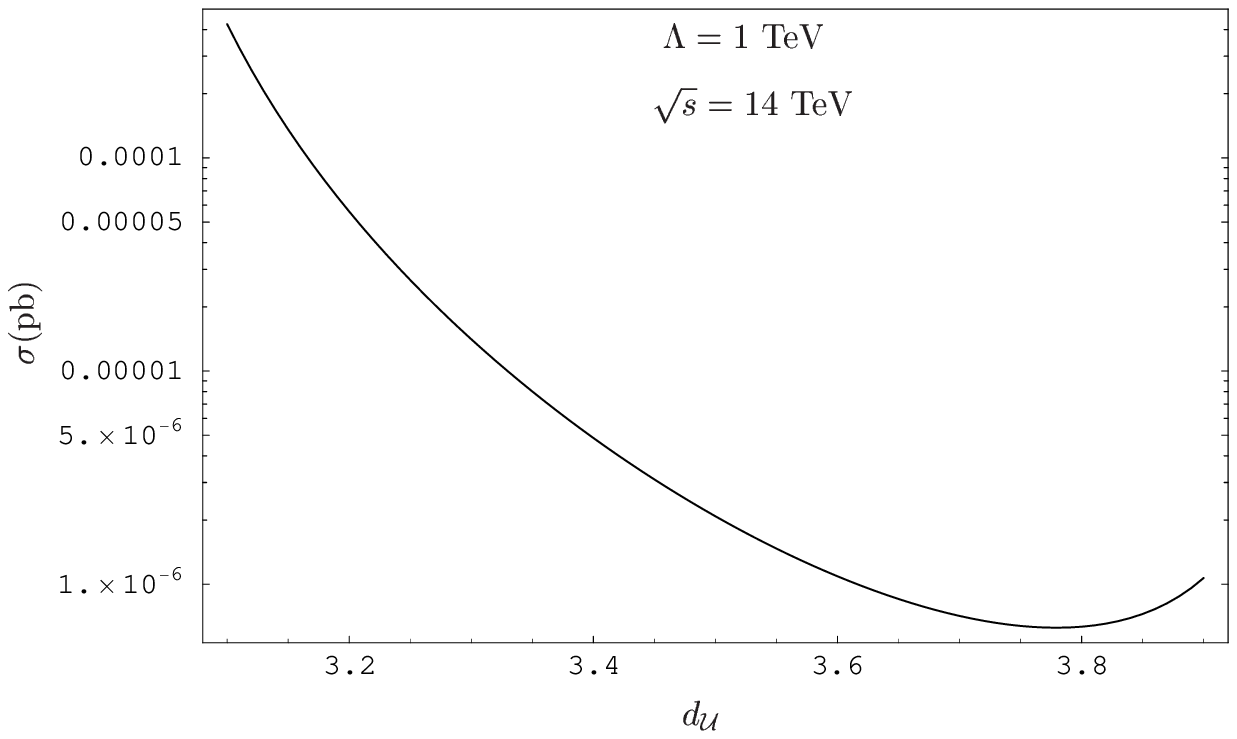}\\
  \caption{Total cross section in pb originating from the reaction $q_i \bar q_j\rightarrow t\bar q_j$ ($q_j=d,b,s$)
   for $\Lambda$= 1 TeV, $\lambda_2=1$,
 at LHC through tensor unparticle}\label{fig6}
\end{figure}

\begin{figure}
  \includegraphics[width=12cm]{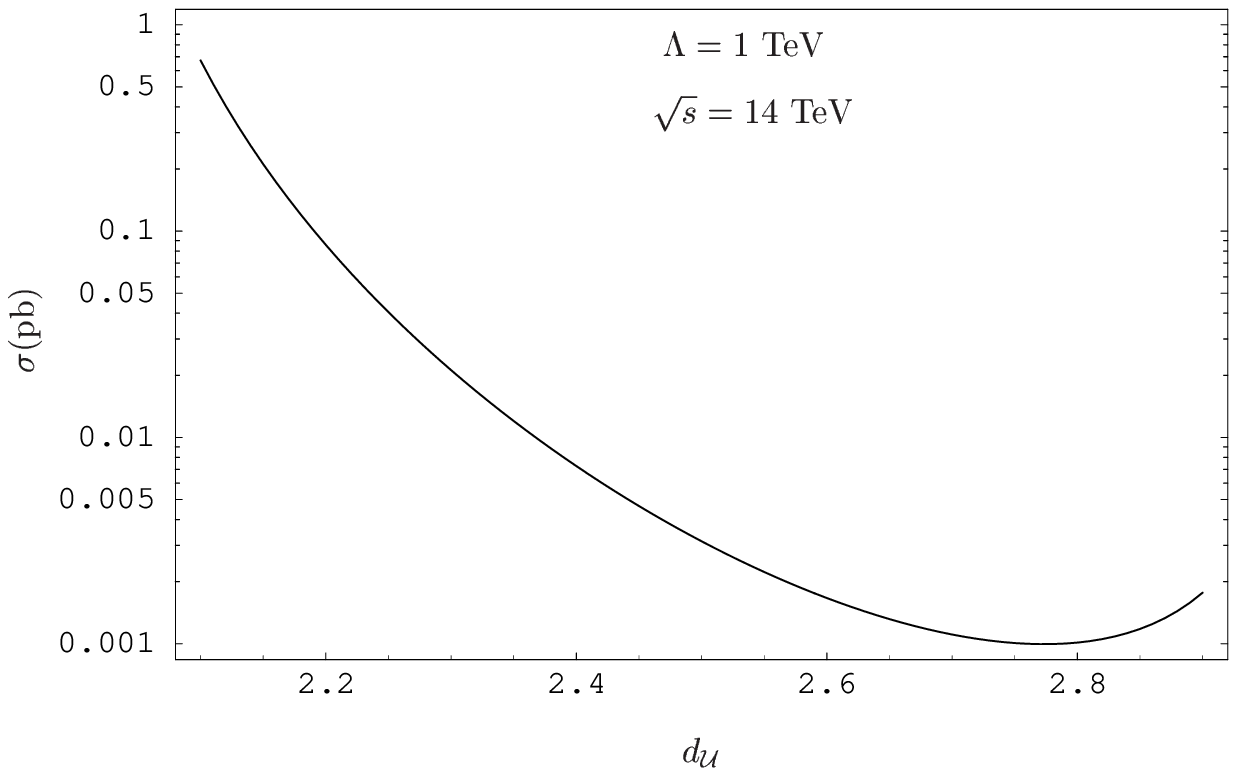}\\
  \caption{Total cross section in pb originating from the reaction $gg\rightarrow t\bar q_i$
  for $\Lambda$= 1 TeV, $\lambda_0=1$ at LHC through scalar unparticle}\label{fig7}
\end{figure}

\begin{figure}
  \includegraphics[width=12cm]{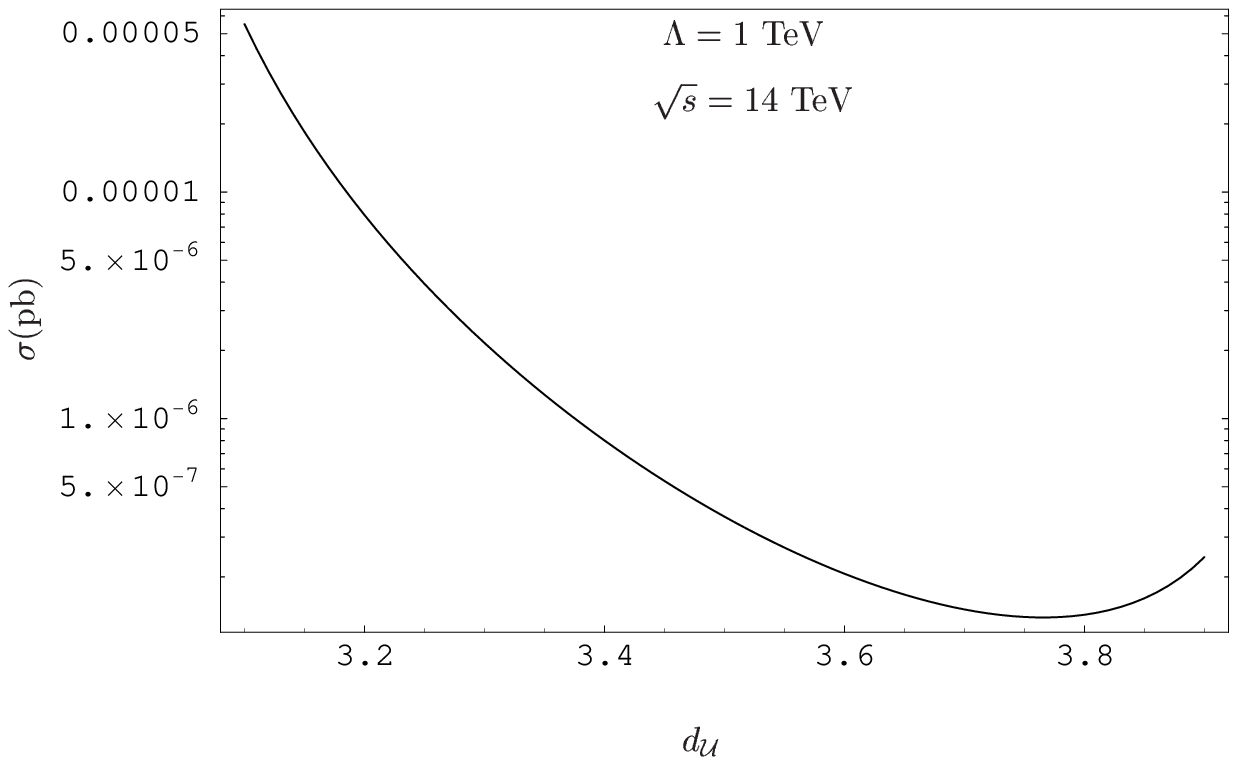}\\
  \caption{Total cross section in pb originating from the reaction $gg\rightarrow t\bar q_i$  for $\Lambda$= 1 TeV, $\lambda_2=1$
 at LHC through tensor unparticle}\label{fig8}
\end{figure}

\begin{figure}
  \includegraphics[width=12cm]{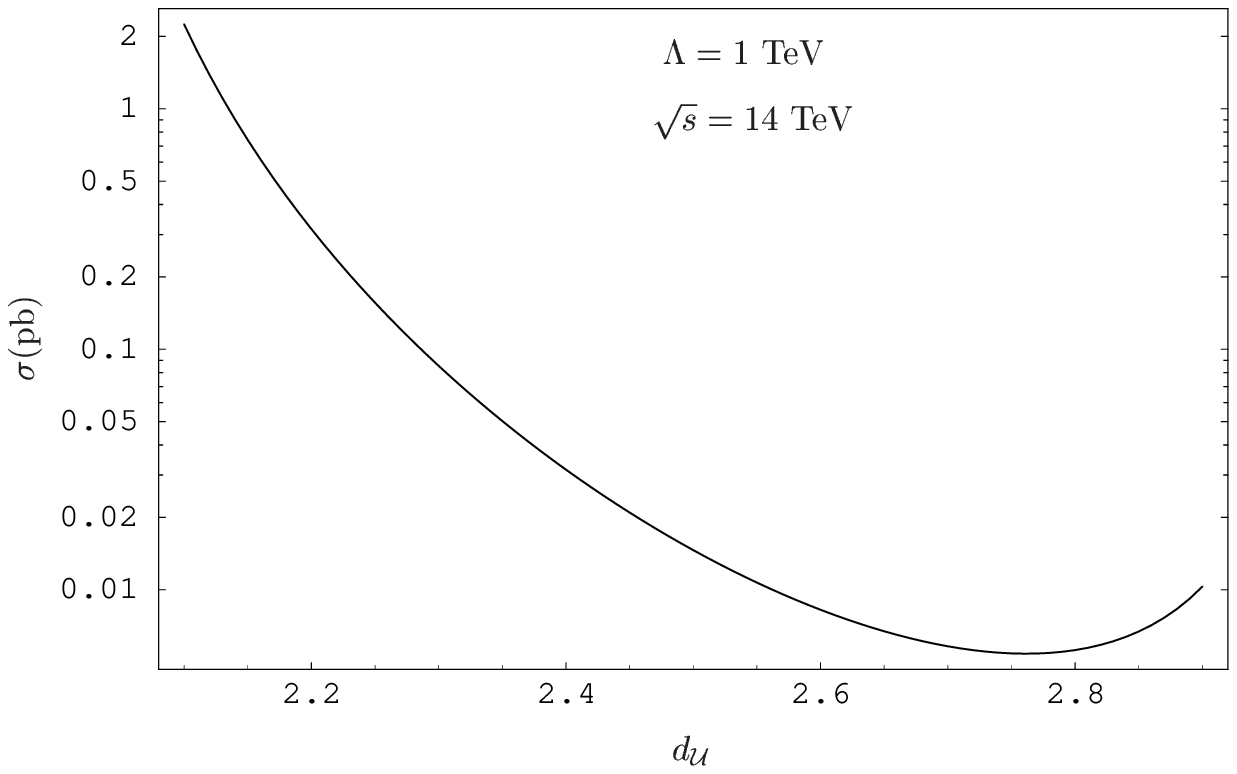}\\
  \caption{Total cross section in pb originating from the reaction $q_ig\rightarrow t g$ for $\Lambda$= 1 TeV, $\lambda_0=1$ at LHC through scalar
unparticle}\label{fig9}
\end{figure}

\begin{figure}
  \includegraphics[width=12cm]{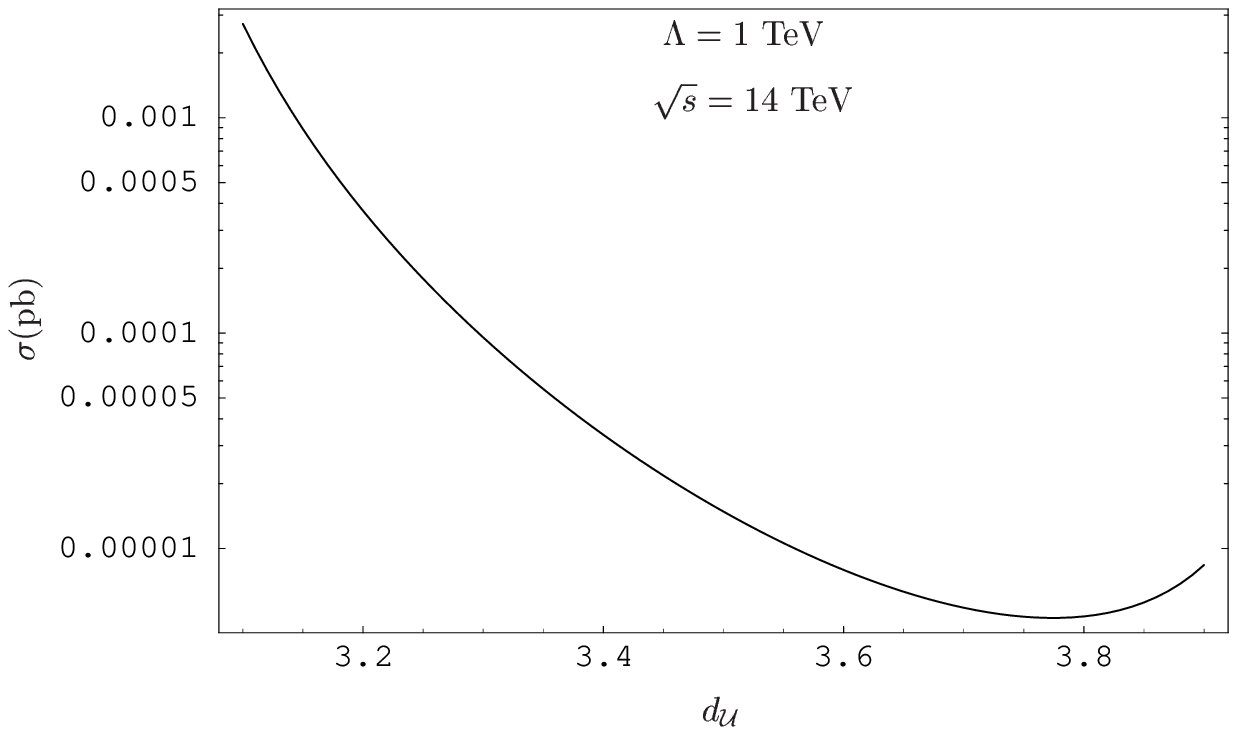}\\
  \caption{Total cross section in pb originating from the reaction $q_ig\rightarrow t g$ for $\Lambda$= 1 TeV, $\lambda_2=1$
 at LHC through tensor unparticle}\label{fig10}
\end{figure}

The LHC precesses are plotted in Figs.~\ref{fig3}-~\ref{fig10}. In
Figs.~\ref{fig3} and~\ref{fig4}, we have plotted, the cross sections
originating from the reactions $pp\rightarrow t\bar q_i+X$ for
$\Lambda$=1 TeV and $\sqrt s$=14 TeV, for vector and scalar, and
tensor cases, respectively. With the large luminosity value of LHC,
$L=10^5$ pb$^{-1}$, we see that about 100 events are possible for
both vector and scalar mediated processes with upper bounds
$d_{\U}=2.48$ and $d_{\U}=2.34$, respectively. For the tensor case
the cross sections are rather small for a wide region of $d_{\U}$,
except at $d_{\U}=3.1$. Here we expect about 30 events per year.

In Figs.~\ref{fig5} and~\ref{fig6} we have plotted the total cross
sections originating  from the t-channel reactions $q_i\bar
q_j\rightarrow t\bar q_j$ ($q_j=d, s, b$) for vector and scalar, and
tensor uparticle contributions, respectively. We expect about 110
events for vector and scalar cases at $d_{\U}=2.38$, and
$d_{\U}=2.34$, respectively. The tensor case is however significant
at $d_{\U}$=3.1, and the expected number of events is about 40.

Figs.~\ref{fig7} and~\ref{fig8} show total cross sections
originating from gluon fusion $gg\rightarrow t\bar q_i$, for the
scalar and tensor cases, respectively. The upper bound on $d_{\U}$
corresponding to hundred events is $d_{\U}=2.75$ for the scalar. The
tensor case, however is rather small with 5 events per year at only
$d_{\U}=3.1$.

Cross sections originating from the reaction $q_ig\rightarrow tg$
are plotted in Figs.~\ref{fig9} and~\ref{fig10} for t-channel scalar
and tensor unparticle mediators, respectively. In the scalar case
the number of signals will be very significant in the interval
$2<d_{\U}<3$ with about 1000 corresponding to $d_{\U}=2.9$. In the
tensor case we get observable number of events, in the interval
$3<d_{\U}<3.14$, with about 100 events per year at $d_{\U}=3.14$.

The cross sections for the associated production of top quarks with
unparticles, $qg\rightarrow t\U$, through s- and u-channels, turn
out to be very small. Namely, we have found that the expected number
of events will be only 7 at $d_{\U}=2.1$ for the scalar case.
Furthermore, the tensor case is seven orders of magnitude more
suppressed than this case. Therefore we are not including any plots
corresponding to these processes.

Finally, our analysis clearly shows that LHC will be the most
suitable platform to probe unparticle physics, the most striking
process being the one originating from $q_ig\rightarrow tg$.

\begin{acknowledgements}
This work is partially supported by Abant Izzet Baysal University
Research Fund.
\end{acknowledgements}

%\newpage

\newpage
%\begin{figure}
 % \includegraphics[width=12cm]{qgtUsca.eps}\\
  %\caption{Total cross section in pb originating from the reaction $qg\rightarrow t \U$ ($q=u,c$)
  %for $\Lambda$= 1 TeV, $\lambda_0=1$ at LHC through scalar unparticle}\label{fig11}
%\end{figure}
%\begin{figure}
%  \includegraphics[width=12cm]{qgtUten.eps}\\
 % \caption{Total cross section in pb originating from the reaction $qg\rightarrow t \U$ ($q=u,c$) for $\Lambda$= 1 TeV, $\lambda_2=1$
 %at LHC through tensor unparticle}\label{fig12}
%\end{figure}
\end{document}